\documentclass[mathleft]{an}
\usepackage{graphicx}
\usepackage{natbib}
\bibpunct{(}{)}{;}{a}{}{,}
\usepackage{longtable}
\usepackage{times}
\bibpunct{(}{)}{;}{a}{}{,}
\begin{document}

% The following seven commands are intended for editorial usage and should be ignored by
% the author(s).
\Pagespan{981}{}% Document's page range. 
% If second parameter is left empty, the last page is computed automatically.
\Yearpublication{2015}%
\Yearsubmission{2015}%
\Month{12}%   
\Volume{336}%  
\Issue{10}% 

\title{Magnetic, chemically peculiar (CP2) stars in the SuperWASP survey}

\author{K. Bernhard\inst{1,2}\fnmsep\thanks{Corresponding author:
  \email{klaus.bernhard@liwest.at, ernham@rz-online.de}\newline}	
\and S.~H{\"u}mmerich\inst{1,2}
\and E. Paunzen\inst{3}}

\institute{Bundesdeutsche Arbeitsgemeinschaft f{\"u}r Ver{\"a}nderliche 
Sterne e.V. (BAV), Berlin, Germany
\and
American Association of Variable Star Observers (AAVSO), Cambridge, USA
\and 
Department of Theoretical Physics and Astrophysics, Masaryk University,
Kotl\'a\v{r}sk\'a 2, 611\,37 Brno, Czech Republic}

\received{2015}
\accepted{2015}
\publonline{later}


\abstract{The magnetic chemically peculiar (CP2) stars of the upper main sequence are well-suited for investigating the 
impact of magnetic fields on the surface layers of stars, which leads to abundance inhomogeneities (spots) 
resulting in photometric variability. The light changes are explained in terms of the oblique rotator model; 
the derived photometric periods thus correlate with the rotational periods of the stars. CP2 stars exhibiting 
this kind of variability are classified as $\alpha^{2}$ Canum Venaticorum (ACV) variables. We have analysed around 
3 850 000 individual photometric WASP measurements of magnetic chemically peculiar (CP2) stars and candidates 
selected from the Catalogue of Ap, HgMn, and Am stars, with the ultimate goal of detecting new ACV variables. 
In total, we found 80 variables, from which 74 are reported here for the first time. The data allowed us to 
establish variability for 23 stars which had been reported as probably constant in the literature before. 
Light curve parameters were obtained for all stars by a least-squares fit with the fundamental sine wave and its first harmonic.
Because of the scarcity of Str{\"o}mgren $uvby\beta$ measurements and the lack of parallax measurements with an accuracy better 
than 20\%, we are not able to give reliable astrophysical parameters for the investigated objects.}
\maketitle

\section{Introduction}\label{introduction}

The group of chemically peculiar (CP) stars on the upper main sequence displays peculiar lines and line strengths, 
in addition to other peculiar features such as a strong global stellar magnetic field \citep{Babc47}. 
One can usually distinguish between Am (CP1), Si as well as SrCrEu (CP2), HgMn (CP3), and He-weak/strong (CP4) stars
\citep{Pres74}. The subgroup of CP2 objects, which comprises B to F-type stars, is characterized by variable 
line strengths and radial velocity changes as well as photometric variability of in general the same periodicity. 
CP2 stars typically have overabundances of up to several dex for Si, Sr, Cr, Eu, and other rare earth elements as 
compared to the Sun \citep{Saff05}; 
the overabundances of the respective elements are strongly correlated with effective temperature.

Photometric variability of the CP2 star $\alpha^{2}$ Canum Venaticorum (ACV) was first reported by \citet{Guth14}.
The light curves of CP stars can be fitted well by a sine wave and its first harmonic with varying amplitudes 
depending on the photometric filter systems \citep{Nort84}. For some CP stars, a double-wave structure of the 
photometric light curves depending on the observed wavelength region is detected \citep{Mait80a}. 
However, similar magnetic field modulus variations are rather rare exceptions \citep{Math97}. 

The variability of CP2 stars is explained in terms of the oblique rotator model \citep{Stib50}, 
according to which the period of the observed variations 
is simply the rotational period. Accurate knowledge of the rotational period and its evolution in 
time is a fundamental step in understanding the complex behaviour of CP2 stars, especially as 
far as it concerns the phase relation between the magnetic, spectral, and light variations \citep{Miku10}. 

\begin{table*}
\caption{Essential data and light curve fit parameters for the 80 stars identified as photometrically variable chemically peculiar stars.}
\label{variables}
\scriptsize{
\begin{center}
\begin{tabular}{lccccccccc}
\hline
\hline 
Star &  Period & Epoch (HJD)       & $A_{\mathrm 1}$ & $A_{\mathrm 2}$   & $\phi_{\mathrm 1}$ & $\phi_{\mathrm 2}$      &  Spectral type &      $(B-V)$ & $(J-K_{\mathrm s})$ \\
     &  [d]            &      [d]            &    [mag]              &  [mag]     & [rad]  &  [rad]    &               & [mag]   & [mag] \\
\hline
HD 653 	&	1.0845(1) 	&	2453978.60(3)	&	0.0034	&	0.0039	&	0.56	&	0.43	&	A0 Cr Eu   	&	+0.111	&	+0.021	\\
HD 7341 	&	1.8815(9) 	&	2454325.68(4)	&	0.0147	&	0.0015	&	0.14	&	0.46	&	A3 Si      	&	+0.003	&	--0.035	\\
HD 8892 	&	1.782(3)  	&	2454323.63(3)	&	0.0179	&	0.0078	&	0.80	&	0.80	&	A0 Si      	&	+0.073	&	--0.034	\\
BD+51 469 	&	1.638(2)  	&	2454332.27(3)	&	0.0038	&	0.0020	&	0.63	&	0.25	&	A Si       	&	+0.043	&	--0.016	\\
BD+36 467 	&	6.221(5)  	&	2453971.5(1) 	&	0.0097	&	0.0015	&	0.04	&	0.96	&	A Sr Eu    	&	+0.300	&	+0.118	\\
HD 15357 	&	2.591(4)  	&	2454348.35(5)	&	0.0085	&	0.0051	&	0.82	&	0.53	&	B9 Si      	&	+0.151	&	+0.018	\\
HD 237040 	&	1.3078(8) 	&	2454350.36(3)	&	0.0530	&	0.0146	&	0.12	&	0.55	&	B9 Si      	&	+0.243	&	+0.070	\\
HD 18410A 	&	5.09(2)   	&	2454347.0(1) 	&	0.0242	&	0.0039	&	0.82	&	0.64	&	A2 Si Cr Eu	&	+0.239	&	+0.050	\\
BD+41 600 	&	5.056(1)  	&	2453197.0(1) 	&	0.0141	&	0.0026	&	0.68	&	0.75	&	A2 Si      	&	+0.076	&	--0.050	\\
BD+49 1011 	&	14.2(1)   	&	2453192.2(3) 	&	0.0067	&	0.0013	&	0.27	&	0.01	&	A0 Cr Eu   	&	+0.165	&	+0.080	\\
HD 279110 	&	0.94622(2)	&	2453217.11(2)	&	0.0063	&	0.0107	&	0.81	&	0.15	&	B9 Si Sr Cr	&	+0.144	&	+0.033	\\
HD 24786 	&	3.0554(8) 	&	2453963.73(9)	&	0.0036	&	0.0001	&	0.45	&	0.20	&	A0 Cr Eu Sr	&	+0.120	&	+0.017	\\
HD 276179 	&	1.2799(3) 	&	2454359.12(2)	&	0.0034	&	0.0076	&	0.67	&	0.05	&	B8 Si      	&	+0.288	&	+0.112	\\
HD 30335 	&	5.100(4)  	&	2454362.9(1) 	&	0.0041	&	0.0055	&	0.17	&	0.50	&	A2 Sr Eu Cr	&	+0.162	&	+0.027	\\
HD 284639 	&	9.136(1)  	&	2453223.1(2) 	&	0.0059	&	0.0039	&	0.00	&	0.47	&	A0 Si Cr   	&	+0.478	&	+0.141	\\
HD 31463 	&	1.445(2)  	&	2454394.55(3)	&	0.0196	&	0.0105	&	0.80	&	0.04	&	B8 Si      	&	+0.000	&	--0.009	\\
HD 34439 	&	1.602(2)  	&	2454397.56(3)	&	0.0258	&	0.0032	&	0.38	&	0.47	&	A Si       	&	+0.062	&	--0.050	\\
HD 280980 	&	12.50(1)  	&	2453221.9(2) 	&	0.0127	&	0.0017	&	0.97	&	0.35	&	B9 Si      	&	+0.240	&	+0.092	\\
HD 281056 	&	34.3(2)   	&	2453269(1)   	&	0.0091	&	0.0019	&	0.87	&	0.40	&	B9 Si Cr Sr	&	+0.116	&	+0.076	\\
HD 243395 	&	1.7605(7) 	&	2453219.86(3)	&	0.0075	&	0.0005	&	0.72	&	0.02	&	A0 Si Sr   	&	+0.297	&	+0.086	\\
HD 243954 	&	3.065(3)  	&	2454145.41(5)	&	0.0035	&	0.0010	&	0.92	&	0.76	&	A1 Si      	&	+0.157	&	+0.026	\\
HD 244531 	&	1.2802(3) 	&	2454017.73(3)	&	0.0030	&	0.0063	&	0.90	&	0.03	&	B9 Si      	&	--0.139	&	+0.151	\\
BD+38 1211 	&	1.3444(3) 	&	2454070.72(3)	&	0.0057	&	0.0033	&	0.47	&	0.40	&	A0 Si      	&	--0.054	&	--0.016	\\
BD+26 859 	&	1.2378(1) 	&	2454011.62(2)	&	0.0090	&	0.0015	&	0.35	&	0.96	&	B8 Si Sr   	&	+0.028	&	+0.188	\\
TYC 2412--87--1 	&	1.1070(5) 	&	2454143.37(2)	&	0.0061	&	0.0042	&	0.23	&	0.36	&	B8 Si Sr   	&	+0.107	&	+0.046	\\
HD 245153 	&	2.854(3)  	&	2454008.73(5)	&	0.0016	&	0.0044	&	0.65	&	0.98	&	A0 Si Sr   	&	+0.286	&	+0.160	\\
HD 245990 	&	1.549(4)  	&	2454069.63(3)	&	0.0110	&	0.0030	&	0.99	&	0.09	&	A1 Si      	&	+0.462	&	+0.180	\\
TYC 2408--1757--1 	&	3.5116(2) 	&	2454031.71(7)	&	0.0105	&	0.0275	&	0.04	&	0.18	&	B9 Si      	&	+0.038	&	+0.034	\\
HD 246993 	&	1.1013(1) 	&	2454122.50(3)	&	0.0079	&	0.0011	&	0.84	&	0.33	&	B9 Si      	&	+0.057	&	+0.062	\\
BD+35 1238 	&	6.680(2)  	&	2454100.6(1) 	&	0.0270	&	0.0010	&	0.41	&	0.35	&	B9 Si Sr   	&	+0.186	&	+0.073	\\
HD 247664 	&	3.034(5)  	&	2454139.45(6)	&	0.0105	&	0.0031	&	0.56	&	0.37	&	B9 Si Sr   	&	+0.148	&	+0.038	\\
HD 247931 	&	14.52(1)  	&	2453269.8(3) 	&	0.0230	&	0.0040	&	0.81	&	0.61	&	B7 Si      	&	+0.167	&	--0.010	\\
HD 248072 	&	1.12172(8)	&	2454145.46(2)	&	0.0070	&	0.0040	&	0.57	&	0.99	&	B9 Si Cr   	&	+0.182	&	+0.026	\\
HD 38943 	&	3.435(1)  	&	2454099.5(1) 	&	0.0093	&	0.0130	&	0.23	&	0.94	&	A Si       	&	+0.123	&	+0.061	\\
HD 248619 	&	1.7823(5) 	&	2454070.52(4)	&	0.0130	&	0.0061	&	0.69	&	0.41	&	B9 Si Sr   	&	+0.127	&	+0.038	\\
HD 248815 	&	4.173(6)  	&	2454148.45(9)	&	0.0126	&	0.0038	&	0.00	&	0.51	&	B9 Si      	&	+0.300	&	+0.108	\\
HD 248769 	&	2.588(3)  	&	2454031.61(6)	&	0.0144	&	0.0012	&	0.98	&	0.14	&	B8 Si Sr   	&	+0.087	&	+0.030	\\
HD 250515 	&	10.58(7)  	&	2454141.5(2) 	&	0.0190	&	0.0043	&	0.30	&	0.92	&	A0 Si      	&	+0.057	&	+0.022	\\
HD 41282 	&	9.33(5)   	&	2454139.5(2) 	&	0.0078	&	0.0012	&	0.34	&	0.80	&	B9 Si      	&	+0.074	&	--0.009	\\
HD 41251 	&	4.0395(5) 	&	2454098.5(1) 	&	0.0250	&	0.0092	&	0.51	&	0.75	&	B9 Si      	&	--0.036	&	--0.004	\\
HD 41844 	&	2.0328(3) 	&	2454454.50(4)	&	0.0069	&	0.0099	&	0.07	&	0.80	&	A0 Si      	&	--0.064	&	--0.096	\\
BD+25 1117 	&	10.98(1)  	&	2454086.0(3) 	&	0.0067	&	0.0004	&	0.66	&	0.42	&	B8 Si      	&	+0.000	&	+0.071	\\
HD 251879 	&	8.434(5)  	&	2454139.5(2) 	&	0.0092	&	0.0019	&	0.95	&	0.54	&	B9 Sr      	&	+0.370	&	--0.006	\\
HD 252106 	&	1.8119(5) 	&	2454135.63(4)	&	0.0108	&	0.0020	&	0.61	&	0.06	&	A0 Si      	&	--0.094	&	+0.007	\\
HD 252104 	&	1.7674(2) 	&	2454085.70(3)	&	0.0058	&	0.0055	&	1.00	&	0.27	&	B9 Si Sr   	&	+0.062	&	+0.069	\\
HD 256008 	&	1.5577(2) 	&	2453240.74(3)	&	0.0126	&	0.0053	&	0.58	&	0.33	&	A2 Sr      	&	+0.248	&	+0.038	\\
CD--35 2960 	&	0.94116(8)	&	2454111.33(2)	&	0.0068	&	0.0072	&	0.75	&	0.94	&	A0 Si      	&	--0.033	&	--0.063	\\
HD 47284 	&	6.854(3)  	&	2454417.6(1) 	&	0.0028	&	0.0096	&	0.92	&	0.68	&	A5 Si Eu Cr	&	+0.432	&	+0.058	\\
HD 49797 	&	1.2263(1) 	&	2454154.37(3)	&	0.0076	&	0.0006	&	0.22	&	0.82	&	B9 Si      	&	--0.111	&	--0.039	\\
HD 268471 	&	1.1295(8) 	&	2453271.70(2)	&	0.0040	&	0.0004	&	0.79	&	0.84	&	A2 Cr Eu   	&	+0.134	&	+0.017	\\
HD 80992 	&	3.381(8)  	&	2454476.60(7)	&	0.0114	&	0.0039	&	0.91	&	0.31	&	A2 Sr Cr   	&	+0.219	&	+0.108	\\
HD 83181 	&	2.5241(3) 	&	2454570.26(5)	&	0.0094	&	0.0066	&	0.12	&	0.43	&	A0 Cr Eu Si	&	+0.036	&	--0.007	\\
HD 93700 	&	3.164(8)  	&	2454142.6(1) 	&	0.0028	&	0.0004	&	0.16	&	0.65	&	A0 Sr Eu   	&	+0.239	&	+0.014	\\
HD 96537 	&	4.305(1)  	&	2454208.37(9)	&	0.0062	&	0.0064	&	0.50	&	0.28	&	A3 Sr Cr Eu	&	+0.288	&	+0.041	\\
HD 107107 	&	9.50(2)   	&	2454537.3(2) 	&	0.0089	&	0.0026	&	0.30	&	0.33	&	A2 Cr Eu Sr	&	+0.179	&	+0.048	\\
HD 107180 	&	10.073(8) 	&	2454580.5(2) 	&	0.0030	&	0.0026	&	0.66	&	0.35	&	A2 Eu Cr Sr	&	+0.413	&	+0.131	\\
HD 109030 	&	0.8537(4) 	&	2454123.76(2)	&	0.0101	&	0.0023	&	0.67	&	0.67	&	A0 Sr      	&	+0.075	&	--0.018	\\
HD 119277 	&	2.774(1)  	&	2454594.40(6)	&	0.0090	&	0.0035	&	0.40	&	0.64	&	B9 Si      	&	--0.085	&	--0.071	\\
HD 128649 	&	2.6014(3) 	&	2454184.53(5)	&	0.0091	&	0.0127	&	0.57	&	0.49	&	A0 Cr Si Eu	&	+0.114	&	+0.070	\\
HD 129102 	&	1.710(3)  	&	2454572.50(3)	&	0.0030	&	0.0070	&	0.35	&	0.70	&	B9 Si      	&	+0.028	&	--0.072	\\
HD 131750 	&	3.318(3)  	&	2454230.65(8)	&	0.0065	&	0.0124	&	0.74	&	0.43	&	A2 Sr Cr Eu	&	+0.267	&	+0.076	\\
HD 131753 	&	1.7583(3) 	&	2453892.34(3)	&	0.0091	&	0.0028	&	0.75	&	0.64	&	B9 Si      	&	--0.038	&	--0.054	\\
HD 135708 	&	2.9644(3) 	&	2453862.35(5)	&	0.0052	&	0.0053	&	0.67	&	0.64	&	A0 Si      	&	--0.023	&	+0.034	\\
HD 142459 	&	1.7272(2) 	&	2453860.39(2)	&	0.0075	&	0.0035	&	0.77	&	0.21	&	A0 Si      	&	+0.188	&	+0.105	\\
HD 143541 	&	0.8887(1) 	&	2454200.07(1)	&	0.0070	&	0.0009	&	0.90	&	0.76	&	A0 Si      	&	+0.453	&	+0.216	\\
HD 148117 	&	2.0579(1) 	&	2454269.11(4)	&	0.0169	&	0.0024	&	0.22	&	0.66	&	A7 Si Eu Cr	&	+0.160	&	+0.194	\\
BD+32 2827 	&	13.236(6) 	&	2453128.5(2) 	&	0.0019	&	0.0033	&	0.88	&	0.60	&	Sr Eu Cr   	&	+0.263	&	--0.009	\\
HD 169887 	&	4.46(1)   	&	2453130.3(1) 	&	0.0135	&	0.0015	&	0.55	&	0.38	&	A0 Si      	&	+0.017	&	+0.020	\\
HD 234924 	&	8.483(5)  	&	2454234.1(1) 	&	0.0232	&	0.0040	&	0.71	&	0.69	&	A2 Sr      	&	--0.045	&	--0.012	\\
HD 195464 	&	2.704(3)  	&	2454272.25(6)	&	0.0102	&	0.0076	&	0.73	&	0.62	&	A0 Si      	&	--0.096	&	--0.116	\\
HD 340768 	&	3.6036(8) 	&	2453129.57(8)	&	0.0230	&	0.0030	&	0.57	&	0.61	&	B9 Si      	&	--0.082	&	--0.087	\\
HD 199187 	&	1.7485(2) 	&	2453862.05(5)	&	0.0043	&	0.0068	&	0.74	&	0.15	&	A0 Si      	&	--0.024	&	--0.047	\\
BD+31 4539A 	&	1.6619(1) 	&	2453993.55(2)	&	0.0320	&	0.0084	&	0.48	&	0.26	&	A0 Si      	&	--0.033	&	--0.008	\\
TYC 3614--2039--1 	&	2.266(3)  	&	2454279.71(6)	&	0.0135	&	0.0083	&	0.80	&	0.73	&	A0 Si      	&	+0.036	&	--0.011	\\
HD 212714 	&	2.9197(3) 	&	2453128.89(6)	&	0.0041	&	0.0118	&	0.02	&	0.46	&	A0 Si      	&	+0.040	&	--0.002	\\
BD+44 4130 	&	2.851(2)  	&	2454290.71(5)	&	0.0279	&	0.0015	&	0.17	&	0.67	&	A0 Si      	&	--0.009	&	--0.002	\\
HD 212899 	&	1.583(2)  	&	2454279.58(3)	&	0.0064	&	0.0019	&	0.99	&	0.50	&	A0 Si      	&	+0.002	&	+0.008	\\
HD 214985 	&	1.3851(1) 	&	2453862.68(3)	&	0.0028	&	0.0004	&	0.60	&	0.67	&	A0 Si      	&	--0.024	&	--0.045	\\
BD+47 4044 	&	13.8(1)   	&	2454297.7(2) 	&	0.0223	&	0.0098	&	0.07	&	0.49	&	B8 Si      	&	+0.133	&	+0.020	\\
BD+46 3957 	&	7.58(2)   	&	2454293.2(1) 	&	0.0100	&	0.0065	&	0.51	&	0.86	&	A Si Sr    	&	+0.167	&	--0.016	\\
\hline
\end{tabular}
\end{center}
}
\end{table*}

Recently, \citet{Balo15} presented an analysis of the light curves of 29 CP1 stars from the Kepler
satellite mission. They found 12 $\delta$ Scuti variables, one $\gamma$ Doradus star and 10 stars, 
whose variability is in accordance with rotational modulation caused by spots.
However, the amplitudes of the detected variability is between 4 and 200\,ppm which, normally, cannot be
achieved via ground based observations. 
This demonstrates that, apparently, even this subgroup of CP stars shows rotationally induced variability.
However, \citet{Auri10} found that none of the 15 investigated A-type stars of peculiarity types other than 
CP2/4 hosts a surface-averaged longitudinal magnetic field of more than 3 Gauss. They concluded that there exists 
a magnetic dichotomy corresponding to a gap of more than one order of magnitude in field strength.

Furthermore, photometric variability among cool-type (F, G, and K) stars due to the presence of starspots 
is a common phenomenon. \citet{Niel13}
published rotational periods of 12\,000 main-sequence stars based on observations by the Kepler
satellite mission. As expected, they found that hot stars (earlier than A-type) rotate faster than their
cooler counterparts. However, it is important to connect all these observations to create a global picture 
of the rotational behaviour of stars.

In this paper, we present the results of the WASP light curve analysis of bona fide CP2 stars and candidates.
Among the 579 investigated objects, we detected 80 ACVs, from which 74 are reported for the first time. 

Observations, target selection, and reductions are described in Sect.~\ref{selection}; data analysis
is characterized in Sect.~\ref{analysis}; 
results are presented and discussed in Sect.~\ref{results}. We 
conclude in Sect.~\ref{conclusion}.

\section{Observations, target selection and reductions} \label{selection}

The main aim of the WASP project is the detection of transiting extrasolar planets. 
Two robotic telescopes are employed, which are situated at the Observatorio del Roque de 
los Muchachos (La Palma) and the South African Astronomical Observatory (SAAO). 
Each telescope consists of an array of eight f/1.8 200mm Canon lenses and 
2048 x 2048 Andor CCD detectors, covering a field of 7.8$\degr$x7.8$\degr$ of sky with an angular size of 
13.7$\arcsec$ / pixel \citep{Poll06}. Initially, observations were taken unfiltered; 
from 2006 onwards, a broadband filter with a passband from $\sim$4000 to $\sim$7000\AA\ was 
employed. Each field was observed about every 9 to 12 minutes \citep{Butt10}. 
The first data release (DR1) of the WASP archive, which encompasses light curve data 
from 2004 to 2008, boasts $\sim$18 million light curves covering a large fraction of the sky 
and provides good photometry for objects in the magnitude range \\ 8\,$\le$\,$V$\,$\le$14\,mag.

A list of targets was established by selecting bona-fide CP2 stars from the 
Catalogue of Ap, HgMn and Am stars \citep[][RM09 hereafter]{Rens09}. The objects
in this catalogue are not explicitly subdivided in the CP groups established by
\citet{Pres74}. We used the listed spectral types therein to distinguish between CP1
stars and the other subgroups (mainly denoted as `Si', `Sr', `Sr Eu Si', `He-weak', `Hg Mn', 
and so on). All objects in this list boasting at least 1 000 data points in WASP DR1 
were investigated. In total, 579 objects were found to meet these criteria.

The corresponding data were downloaded from the WASP archive at 
the computing and storage facilities of the CERIT Scientific Cloud\footnote{http://wasp.cerit-sc.cz/} \citep{Paun14}. 
To avoid the most significant saturation effects, all objects brighter than \\ $V$\,$<$\,8\,mag were eliminated. 
A lower magnitude cut-off was not deemed necessary as there are only 40 objects in the 
RM09 catalogue with $V$\,$>$\,14\,mag.

\section{Data Analysis} \label{analysis}

As a first step, all light curves were inspected visually and obvious outliers and data points associated to exceedingly 
large error bars were removed. The data were then searched for periodic signals in the frequency domain of 0\,$<$\,f (c/d)\,$<$\,50 
using \textsc{Period04} \citep{Lenz04}. Objects exhibiting periodic signals well above the noise level (corresponding to a 
semi-amplitude of at least $\sim$0.005\,mag, as determined with \textsc{Period04}) were subjected to a more detailed analysis.
In this second step, the data were binned in order to increase the accuracy of the measurements. Depending on the length 
of the dataset, the number and cadence of observations and the quality of the data, different bin-sizes from 0.005 -- 0.05\,d 
were chosen. The data were then carefully cleaned from remaining systematic trends, which were mostly of instrumental 
origin or due to blending issues. In addition to that, the data of some stars in the magnitude range 8\,$<$\,$V$\,$<$\,9 mag were 
also found to suffer from significant systematic trends likely due to saturation effects \citep[cf.][]{Smal14}. In some 
cases, the severity of artifacts present in the data necessitated the removal of entire epochs or the complete dataset 
of one of the WASP cameras.

WASP observations are often made up of distinctive parts separated by observational gaps of varying length in the data. 
Where this applied, shifts in mean magnitude between the corresponding parts of the data were sometimes observed. The 
light curves were consequently detrended by shifting all parts of the data to the mean magnitude of the combined dataset.
In some cases, systematic trends introduced spurious periods in the data and might have effectively masked any low-amplitude 
variability present. Stars exhibiting a weak signal that could not be attributed to systematic trends but did not produce a 
convincing phase plot either, were generally rejected in order to keep the sample free of possibly spurious detections that 
might contaminate the sample of derived rotational periods.

The data were searched for periodic signals using the Phase Dispersion Method (PDM) of \citet{Stel78} and the Analysis of 
Variance (ANOVA) statistic developed by \citet{Schw96}, as implemented in the \textsc{PERANSO} software package \citep{Paun15}. 
Periods were searched in the range of 0.1\,$<$\,P (d)\,$<$\,50. The resulting 
power spectra were examined for significant features, and the data were folded with the resulting best-fitting periods and visually inspected.
Objects exhibiting convincing phase plots were considered for inclusion in the final sample. The General Catalogue of Variable Stars \citep[GCVS, ][]{Samu07}, 
the AAVSO International Variable Star Index, VSX \citep{Wats06}, the VizieR \citep{Ochs00}, and SIMBAD \citep{Weng00} databases were 
consulted to check for an entry in variability catalogues. Objects that have already been announced as ACV variables in the 
literature were dropped from our sample, the only exception being V499 Per, for which only a tentative period has been published in the literature.

It has been shown in the literature that, in most cases, the light curves of CP2 stars can be well represented by a sine wave and 
its first harmonic \citep[e.g.][]{Nort84, Math85, Heck87}. A least-squares fit to the observations was done using the program 
package \textsc{Period04}. Each light curve was fitted using a Fourier series consisting of the fundamental sine wave and its first harmonic, 
from which the corresponding amplitudes and their phases were derived. The light curve parameters 
($A_{\mathrm 1}$, $A_{\mathrm 2}$, $\phi_{\mathrm 1}$, and $\phi_{\mathrm 2}$) are listed in Table \ref{variables}.

The object was finally classified according to spectral type, colour information, period and shape of the light curve. The observed 
variability pattern of all stars in the present sample is in accordance with rotational modulation caused by spots. 
For some few cases, the discrimination between the light curves of double-waved ACVs and the variability induced by orbital motion 
(ellipsoidal variables/eclipsing variables) is not straightforward. As the incidence of ellipsoidal or eclipsing variables among CP2 stars 
is very low \citep{Gerb85,Nort04,Hubr14,Bern15}, we are inclined to interpret the observed variability as being due to rotational modulation.
The initial period search with \textsc{Period04} in the frequency range up to 50 c/d also resulted in the discovery of six new $\delta$ Scuti 
variables. These will be dealt with in an upcoming paper.

Table \ref{variables} lists the results and some additional observational data. It is organised as follows:

\begin{itemize}
\item Column 1: Star name, HD number, or other conventional identification
\item Column 2: Period (d)
\item Column 3: Epoch (HJD), time of maximum
\item Column 4: Amplitude of the fundamental variation ($A_{\mathrm 1}$)
\item Column 5: Amplitude of the first harmonic variation ($A_{\mathrm 2}$)
\item Column 6: Phase of the fundamental variation ($\phi_{\mathrm 1}$)
\item Column 7: Phase of the first harmonic variation ($\phi_{\mathrm 2}$)
\item Column 8: Spectral classification, as listed in RM09
\item Column 9: $(B-V)$ index, taken from \citet{Khar01}
\item Column 10: $(J-K_{\mathrm s})$ index, as derived from the 2MASS catalogue \citep{Skru06}.
\end{itemize}

In agreement with the findings of other investigators, we confirm that WASP data are very well suited to investigate variable stars
with low photometric amplitudes \citep[cf.][]{Smal14}.

\begin{figure*}
\begin{center}
\includegraphics[width=1.0\textwidth]{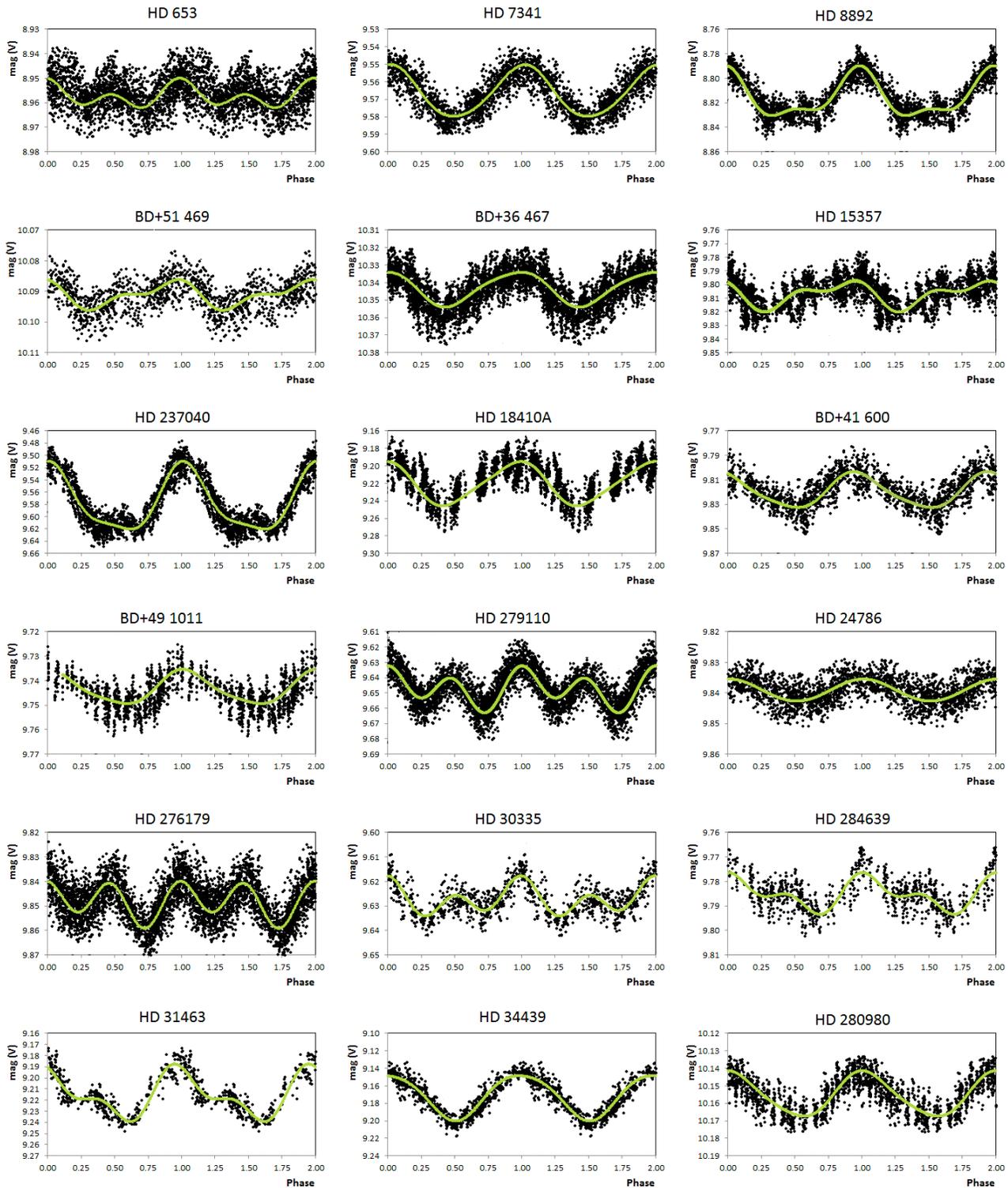}
\caption{The light curves of all objects, folded with the period listed in Table \ref{variables}. The fit curves 
corresponding to the light curve parameters given in Table \ref{variables} are indicated by the solid lines.}
\end{center}
\end{figure*}

\addtocounter{figure}{-1}

\begin{figure*}
\begin{center}
\includegraphics[width=1.0\textwidth]{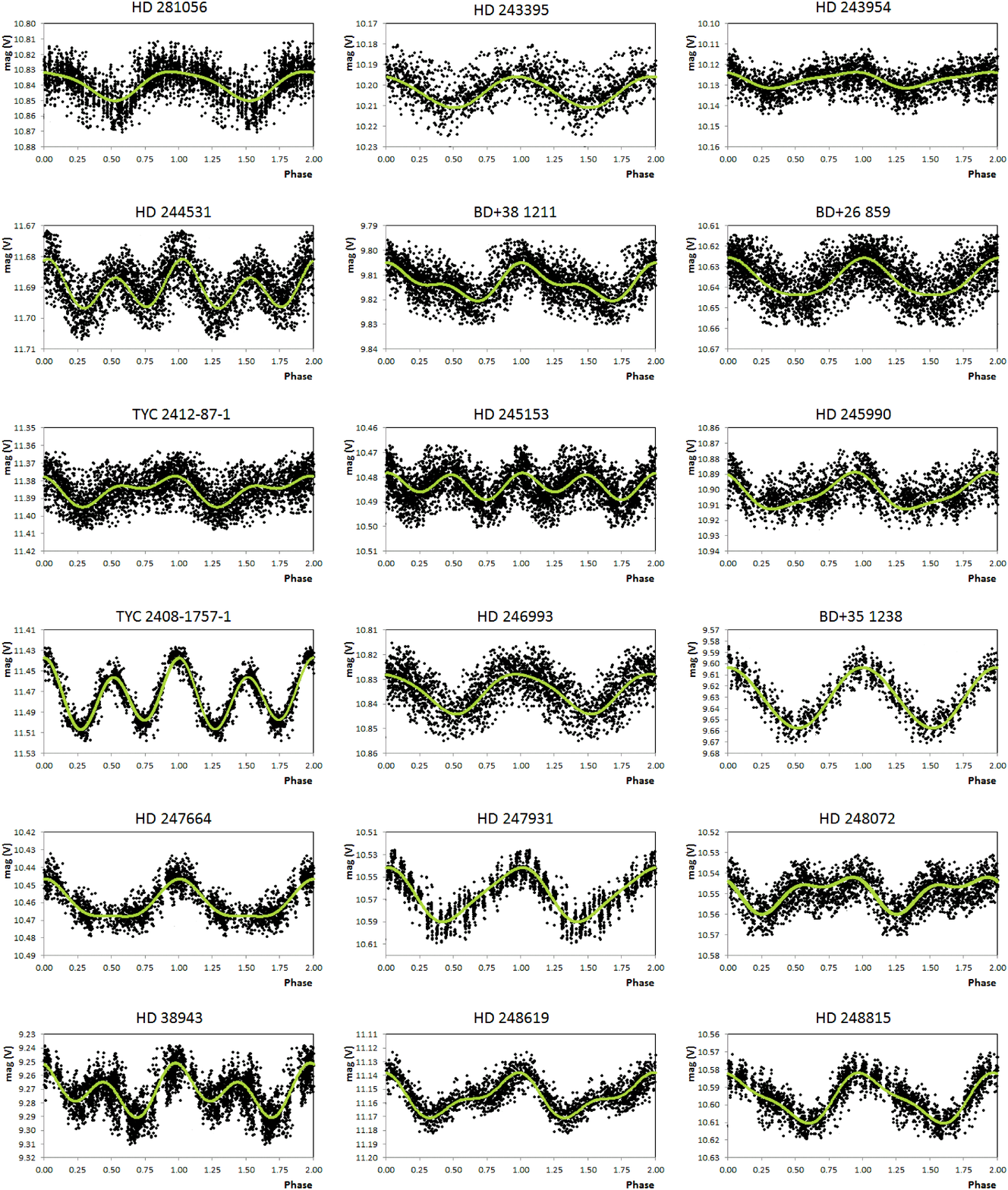}
\caption{continued.} 
\end{center}
\end{figure*}

\addtocounter{figure}{-1}

\begin{figure*}
\begin{center}
\includegraphics[width=1.0\textwidth]{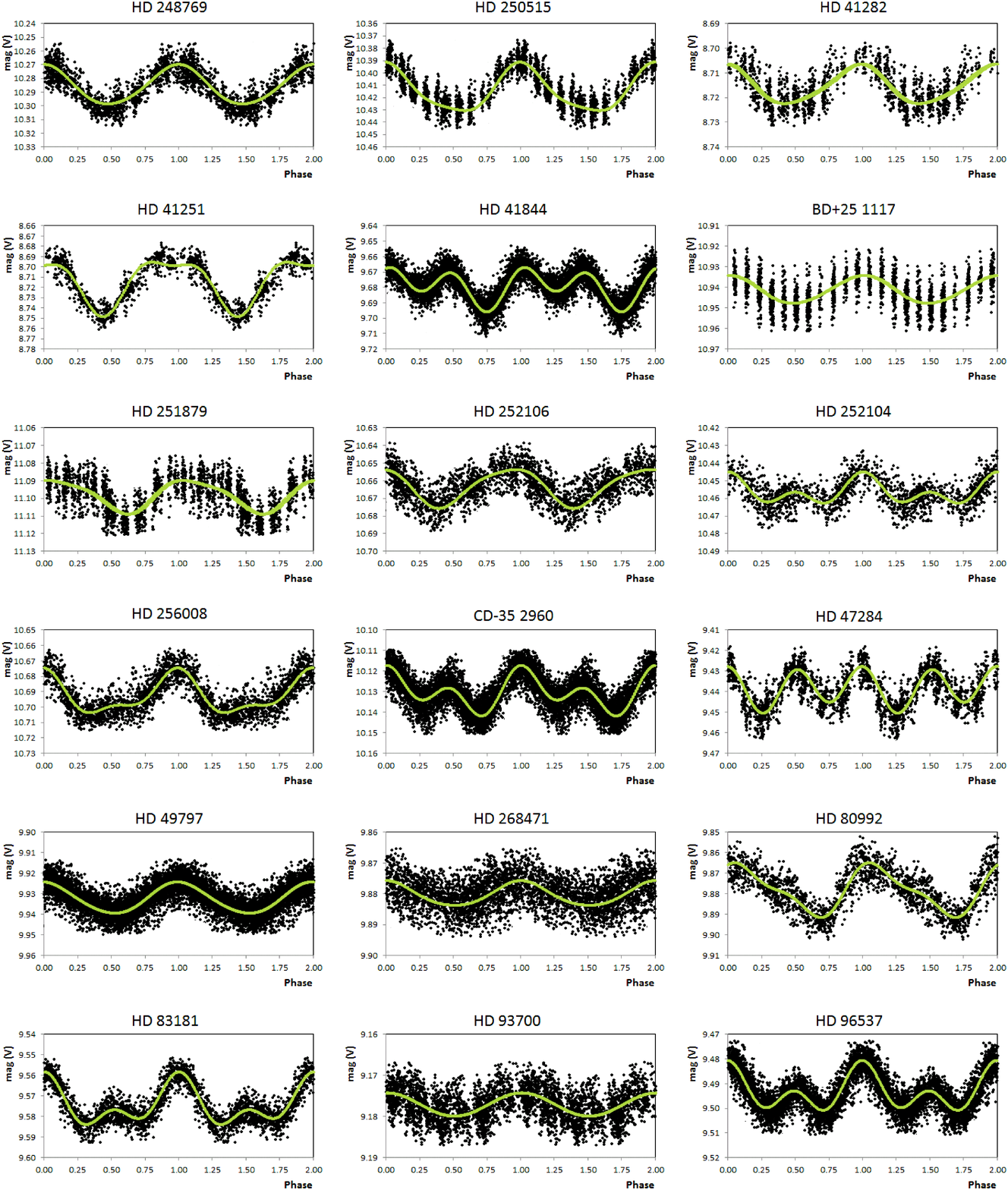}
\caption{continued.} 
\end{center}
\end{figure*}

\addtocounter{figure}{-1}

\begin{figure*}
\begin{center}
\includegraphics[width=1.0\textwidth]{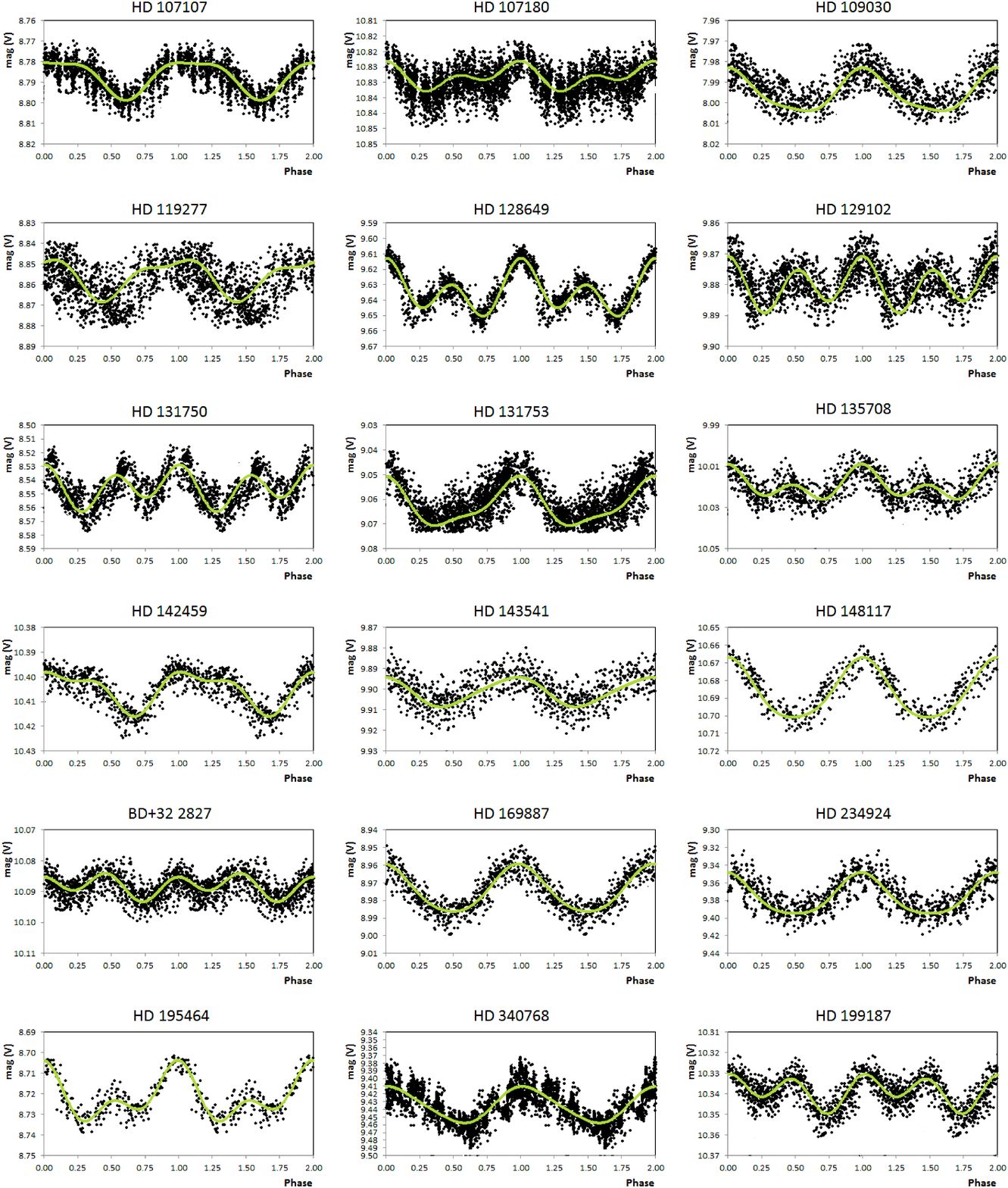}
\caption{continued.} 
\end{center}
\end{figure*}

\addtocounter{figure}{-1}

\begin{figure*}
\begin{center}
\includegraphics[width=1.0\textwidth]{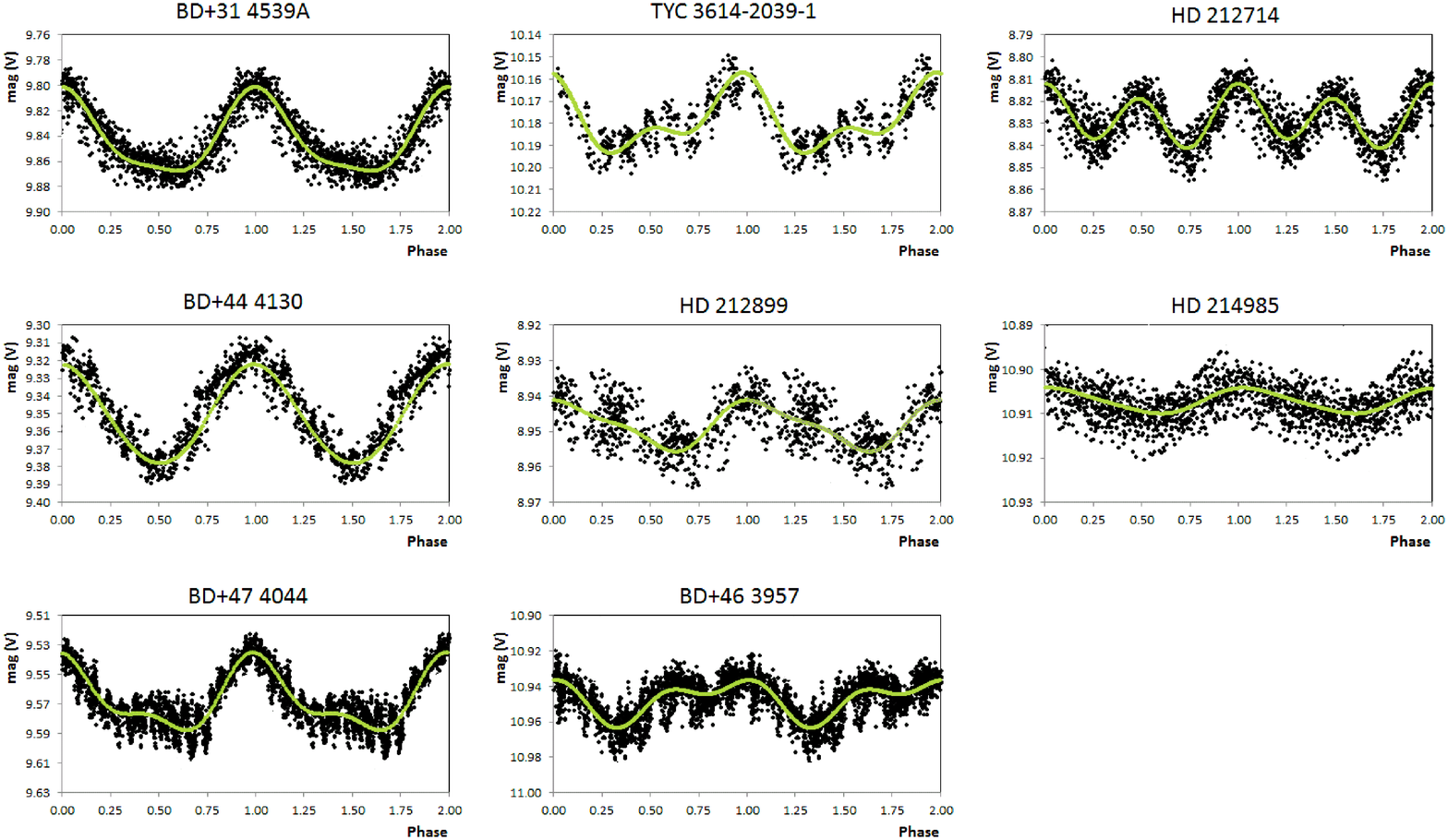}
\caption{continued.} 
\end{center}
\end{figure*}

\section{Results} \label{results}

The following stars have already been studied in the past and discussed in more detail.

{\it HD 8892:} \citet{Koen02} published a period of 1.77945\,d based on Hipparcos photometry. Within the
errors, this is in line with our result. \citet{Rimo12} classified it as slowly pulsating B-type star 
using automatic classification based on Bayesian networks (probability of 0.58) and the prediction by random forests (probability of 0.33).
The spectral classification of Ap Si \citep{Cowl65} is consistent with its colours. We are therefore confident that this is a
true CP2 and ACV object.

{\it HD 18410A:} The period of 5.08053\,d listed by \citet{Koen02} is in agreement with the one derived by us. 
\citet{Rimo12} list a variable type of either `BE + GCAS' or ACV depending on the used classification
method. 

{\it HD 41251:} V448 Aur, there are two different variability types found in the literature. The VSX
catalogue lists it as slowly pulsating B-type star whereas \citet{Duba11} list an ACV type. The given
periods are comparable to ours.

{\it HD 131750:} \citet{Stro66} reported variability for this star (NSV 6859). They do not list a period but
a photographic amplitude of 0.35\,mag, which - originating in an early type object - is rather large if due 
to rotation and/or pulsation. In the same paper, three additional A-type stars of comparable amplitudes are 
presented (HD 148891, HD 188297, and HD 204370), which are all eclipsing binary systems. However, our data 
show no eclipses, which is also supported by an analysis of Hipparcos data. In addition, the search for roAp 
characteristics gave a null result \citep{Frey08}. 

{\it HD 250515:} This star is listed in the ASAS Catalogue of Variable Stars \citep[ASAS,][]{Pojm05} as a 
'MISC' type object with a period of 0.47683 d. \citet{Wrai12} found no variability whereas 
\citet{Rich12} list a period of 0.9116\,d and an ACV type (probability of 0.4). The period that produces the best fit to the 
WASP data (10.58\,d) is much longer than the ones that have been reported before. In order to investigate this issue, we have 
combined WASP and ASAS data, with the latter dataset consisting of only 47 datapoints. The 10.58\,d period produces the best 
fit to the combined data.

{\it HD 279110:} This star (V499 Per) is located in the Perseus OB1 association. The period analysis by \citet{Nort87}
was not conclusive (possible period of 0.48 or 0.96\,d) due to the small number of available observations. From
the WASP data, which have an excellent spatial coverage, we can definitely conclude that 0.94622\,d is the correct 
period.

{\it HD 284639, HD 243395, HD 243954, HD 244531, BD+26 859, HD 245153,  HD 245990, TYC 2408-1757-1, 
HD 246993, HD 247664, HD 247931, HD 248072, HD 248619, HD 248769,  HD 248815, HD 41282, BD+25 1117, 
HD 251879, HD 252106, HD 252104, HD 256008, HD 268471, and HD 148117:} These 23 stars were reported as constant or 
probably constant by \citet{Wrai12}, who analysed the light curves of 337 probable magnetic chemically peculiar 
stars obtained with the STEREO spacecraft. 82 stars were detected as variable, of which 48 were reported for the first time.
The authors stated that the stars, which are classified as constant, might be intrinsically variable if their 
period is, for example, longer than 10\,d or their light curve is affected by substantial blending or
systematic effects. From the sample, only five stars have periods close or above 10\,d (Table \ref{variables}): 
HD 284639 (9.136\,d), HD 247931 (14.52\,d), HD 41282 (9.33\,d), BD+25 1117 (10.98\,d), HD 251879 (8.434\,d).

In accordance with the findings of previous investigators \citep[e.g.][]{Nort84, Math85, Heck87}, 
our light curve analysis confirms that the light curves of most CP2 stars can be adequately described by a sine wave 
and its first harmonic (cf. Section 3 and Fig. 1.)

For an astrophysical analysis of a possible correlation of the found rotational
periods with age and mass, for example, the location of the individual stars in the
Hertzsprung-Russell-diagram (HRD) has to be estimated. For this, the 
effective temperature (or colour) and luminosity (or absolute magnitude) need to be calibrated.
Such calibrations have been especially developed and tested for the different CP subgroups \citep{Mait80b,Neto08}
but they all require some additional information such as photometric data in various systems, 
an estimation of the reddening, and, most important, knowledge of the distance.

For all stars, Johnson $BV$ \citep{Khar01} and 2MASS \citep{Skru06} colours are available. The errors
for $(B-V)$ are between 0.01 and 0.2\,mag whereas they are between 0.02 and 0.05\,mag for $(J-H)$.
For only four stars (HD 30335, HD 109030, HD 1237040, and HIP 109911), the complete Str{\"o}mgren $uvby\beta$
measurements are available. 

With these sparse data, it is not possible to calibrate the reddening of our sample stars. 
We therefore used the Galactic location of our targets to calculate the maximum reddening using the 
model by \citet{Schl11}, who obtained reddening as the difference between the measured and predicted 
colours of a star as derived from stellar parameters from the Sloan Digital Sky Survey.
For our sample, we find a maximum of $E(B-V)$\,=\,1.92\,mag, with a mean of 0.43\,mag and a median of
0.33\,mag, respectively. Bearing in mind that the total absorption $A_{\mathrm V}$\,=\,3.1$E(B-V)$, the 
importance of a reasonable determination is obvious. For none of our targets, a parallax measurement
with an accuracy better than 20\% is available \citep{Leeu07}. Taking into account the above listed limitations,
we are not able to give reliable astrophysical parameters for the investigated objects.

\section{Conclusion} \label{conclusion}

We have carried out a search for photometric variability in confirmed or suspected CP2 stars from the Catalogue of 
Ap, HgMn, and Am stars (RM09) using the publicly available observations from WASP DR1. Around 3 850 000 individual 
photometric measurements were analysed, which resulted in the discovery of 80 variables, from which 74 are reported 
here for the first time. Among this number are 23 stars which had been reported as probably constant in the literature 
before. In agreement with the literature, our light curve analysis confirms that the light curves of most CP2 stars 
can be adequately described by a sine wave and its first harmonic.
Because of the scarcity of suitable photometric data and the lack of parallax measurements with an accuracy 
better than 20\%, we are not able to give reliable astrophysical parameters for the investigated objects.

\begin{acknowledgements}
The WASP project is funded and maintained
by Queen's University Belfast, the Universities of Keele, St.
Andrews, Warwick and Leicester, the Open University, the Isaac
Newton Group, the Instituto de Astrofisica Canarias, the South
African Astronomical Observatory and by the STFC. This project was supported by the SoMoPro II Programme (3SGA5916),
co-financed by the European Union and the South Moravian Region, the 
grant GA \v{C}R 7AMB12AT003, LH14300, and
the financial contributions of the Austrian Agency for International 
Cooperation in Education and Research (BG-03/2013 and CZ-09/2014).
This work reflects the opinion of the authors and the European 
Union is not responsible for any possible application of the information
included in the paper.
\end{acknowledgements}

\end{document}